\documentclass{ws-procs961x669}            % default, citations in superscript
\begin{document}

\title{Summary of the parallel session HR3}
%Summary of the session, Time and Philosophy in Physics
\author{Shokoufe Faraji}

\address{%
 University of Bremen, Center of Applied Space Technology and Microgravity (ZARM), 28359 Germany
}%    

\begin{abstract}
This is the summary of the parallel session entitled "Time and Philosophy in Physics", chaired by Shokoufe Faraji in the sixteenth Marcel Grossmann Meeting. This parallel session aimed to discuss open issues related to Time and fundamental laws from different perspectives in a complementary point of view.
\end{abstract}

\keywords{Time in Physics, Time in Philosophy; passage of Time; fundamental laws}

\bodymatter

\section{Introduction}
It is astonishing that we can understand the Universe in the way we study it. However, concurrently celebrating the achievements of science, we should respect its limits and not claim more than it can actually achieve or attain, where by pushing its limits, complex philosophical issues may arise. However, the concern is about when theories are applied beyond their range of applicability. Since the visible Universe may not give us enough information to characterize the laws of physics completely, thus we need a more comprehensive framework that not only incorporates science but can also go beyond the limits.

Therefore, for example the questions on the domain of validity of the fundamental laws of physics and the nature of existence of infinities are sound viable. However, maybe above all, the nature of Time and the order present in nature is the greatest secret in the Universe \cite{penrose-roadtoreality-2005,ellis2006ebu,auletta2008top,article,ellistopdown}. 
%\cite{ellis1972global,ellis2002cosmology}
% and this is time to reconsider questions like whether philosophy plays an essential role in the development of science, or is more damaging than helpful for physics, "because the big questions that used to be discussed by philosophers are now in the hands of physicists".

\section{The passage of Time and top-down causation}

This is reported by Prof. Barbara Drossel.

There is a common underlying agreement among many physicists to consider the properties of the fundamental equations of physics are also properties of nature. Therefore, since these equations are deterministic, it is concluded that nature is also deterministic. Besides, the Time evolution is completely determined by the initial state and the fundamental laws. Furthermore, these laws are invariant under changing the direction of Time. However, any physical theory is an approximation and not an exact description of the physical processes of interest. And by no means it is a complete description of everything that is going on in the physical world. Even a deterministic and Time-reversible fundamental theory is in practice supplemented by irreversible and stochastic features. In conclusion, the irreversibility and stochasticity enter when the influence of the environment on a system is taken into account, which is related to emergence and top-down causation and microsystems are causally open to influences from the macroscopic environment.

\section{Temporal asymmetry and causality} 

This is reported by Prof. Wayne Myrvold.

The temporal asymmetry between past and future permeates virtually every aspect of the world of our experience. As far as we know, it has no counterpart in the laws of fundamental physics. One option is to trace this asymmetry to a fact about the early state of the universe, either taken as a contingent,  unexplained fact, or as a consequence of some physical principle. However, there is an alternate route to explaining temporally asymmetric phenomena, according to how work on the process of equilibration is done. This does not involve any hypothesis about the past, rather it is founded on conditions that we can apply at any Time, closely related to the concept of causality \cite{book}. In fact, temporal asymmetry enters our explanations because the very notion of what it means to explain something is temporal asymmetry. To sum up, the conceptions of casual order and temporal order are not independent and in fact they are two sides of the same coin. 

Therefore, the absence of temporally asymmetry in the fundamental laws is no obstacle to explaining that asymmetry. What is invoked is a temporal asymmetry in the very notion of explanation.

\section{Temporal ordering in Philosophy and Physics}

This is reported by Prof. Norman Sieroka.

Time and temporal order are one of the oldest and most eminent concepts in both philosophy and physics.
One of the main aims of philosophy is to coordinate the human experience in structural analysis and comparison. For this, one needs to acknowledge the differences in types of Time and Time scales. Then we realize many issues about Time are issues about the ordering of events both within and across different types and scales \cite{articles}. 

From the philosophical point of view, mainly there are two different types of temporal orderings. The first one, the presentism or possibilism perspective; events being characterized by the present, past or future, and try to focus on mental states and experience; however, this view may have trouble in physics if simultaneity is not absolute. The second view is Eternalism; the ordering relation is only about events being earlier or later and focuses on physical states, and there is no vital distinction between past, present and future. However, it seems the second one to be of importance to physics. 

To sum up, the relation between these two orderings and their inner dynamics can be interesting in the context of the interaction between philosophy and physics. Besides, investigation of the notion of an \textit{event} seems fundamentally essential in both areas.

\section{Time in Quantum Gravity: a false problem}

This is reported by Prof. Carlo Rovelli.

There are several problems of Time in Physics that are related to each other. Indeed, always a lot of confusion comes from mixing problems that are distinct one from another. In particular, an extensive discussion of Time in Quantum Gravity was caused in the 60th by the Wheeler-de Witt equation, which at first sight suggests Time is frozen and there is no evolution. However, dealing with Time in General Relativity is a slightly different than in non-relativistic Physics \cite{PhysRevD.43.442}. The difficulty for developing Time in General Relativity is to struggle with the meaning of the coordinates. In fact, coordinates in non-relativistic field Theory are measurable by rods and clocks; in contrast to General Relativity, there is nothing physical attached to coordinates. In fact, all possible versions of Time in General Relativity is a quantity that is locally dependent on the Gravitational field. This means the entire formulation of dynamics as evolution in Time does not fit within the scope of General Relativity. In fact, instead of conceptualizing the evolution of physical variables evolving with respect to a variable, we are forced to consider the evolution of a variable with respect to one another. So by this perception, the relativistic dynamics is not frozen in Quantum Gravity and simply described in a different language. Thus, we can predict the relation among a number of variables, and there is no mystery about \textit{emergence of Time} in Quantum Gravity. In fact, many tentative quantum gravity theories, from the Wheeler-de Witt equation to Loop Quantum Gravity, do not specify a Time variable, and yet they are predictive. Of course, this view has substantial consequences, for example, there are some indications that the Time might be discrete.

\section{The issue of Time in Fundamental Physics}

This is reported by Prof. Abhay Ashtekar.

%Dynamics are govern by constrain.

%Provided a global perspective on the evolutions of our understanding of the issue of Time. 
 
The Notion of Time in fundamental physics has undergone radical revisions over centuries dramatically \cite{2013arXiv1312.6322A}. However, new observational windows continually open up in the process. In particular, issues at the forefront of cosmology and quantum gravity pose new conceptual and technical challenges. The same is true in particle Physics, GPS systems or Quantum Information Theory.

A pleasing feature is coming together from physical and operational ideas, fundamental concepts from General Relativity, the associated Hamiltonian constraint, and a post-Pauli understanding of how one should think of Time in Quantum Mechanics. In fact, in Quantum Mechanics, the focus is on the extended system by considering the clock. In Quantum Gravity, there is no such thing as Time in Quantum mechanics, and the extended system kinematics is already contained the dynamical variables of General Relativity.

However, it does not mean \textit{everything is understood and well-controlled}. Many refinements, extensions and technical issues are still open; however, there is no conceptual roadblock to this understanding.

\section{Time in General Relativity and Quantum Mechanics}

This is reported by Prof. Claus Laemmerzahl.

The nature of Time is characteristic of the underlying theory. Within the formalism of General Relativity, it is possible to define or characterize a distinguished clock and Time operationally, and this is the so-called standard clock providing proper Time. In addition, within Quantum Mechanics, it is possible to define a clock and the corresponding Time. This clock is an atomic clock, and it provides Time in the unit of the second. It can be proven with high precision that both clocks are compatible though they are based on entirely different notions \cite{etde_5613804}. Based on such standard clocks, we define Special Relativity effects such as the gravitational redshift, gravitational Time delay and gravitomagnetic clock effect. This compatibility breaks down in strong gravitational fields and also in generalized theories of gravity.

It would be of some interest to study the relativistic two-body problem and another Time scale from chirp. In addition, studying a bound system in an external gravitational field like the Earth-Moon system in the field of the Sun, seems necessary to our understanding.

\section{Time's passage}

This is reported by Janathan Dickau.

The way we experience Time is in the accumulation of experiences and events that happen in the moment. On the one hand, philosophers have tried to explain both the nature of Time and its origin. On the other hand, the meaning of Time in different areas of physics like in Classical and Quantum Mechanics, and Relativity are different. So we hope for example Quantum Gravity theories will help resolve this issue. Recent advances in Mathematics also hold promise for a unified basis explaining both the thermodynamic and quantum-mechanical Time arrows in a consistent way. It is important to continue our exploration of quantum gravity theories and try to understand the cosmological context for these theories in a broader way. So physicists should see it as a responsibility to build bridges between the philosophically disconnected islands of thought, by applying a Philosophy of Physics that supports our common endeavor to learn the secrets of the cosmos.

\section{A glimpse of Feynman's contributions to the debate on the Foundations of Quantum Mechanics}

This is reported by Dr. Adele Naddeo.

One of the most debated questions in the famous 1957 Chapel Hill conference on “The Role of Gravitation in Physics” was whether or not the gravitational field has to be quantized \cite{Dewitt2011TheRO}. Feynman proposed several gedanken experiments in order to argue in favor of the necessity of gravitational quantization, and hinting to decoherence as a viable solution to the problem of wave function collapse \cite{2021EPJH...46...22D}. Feynman believed that nature cannot be half classical and half quantum; therefore, we had the unified field theories. After quantum theory one tries to quantize gravity!

Another issue was related to the limitations that quantum quantum theory may impose on the measurements of space-time distances and curvature. However, since the Planck mass does not set a lower limit to the mass of a particle whose gravitational field can be measured, so the gravitational field of any mass can in principle be measured.

%Finally, concerning the role of the observer in a closed Universe. 

Feynman also discussed the role of the observer in quantum
mechanics. The Schr\"{o}dinger's cat paradox allowed him to
illustrate the difference between the results of a measurement
carried out by an external as well as an internal observer \cite{2011EPJH...36...63Z}. While the external observer describes his results by an amplitude, according to the internal observer the results of the
same measurement are given by a probability. In other words, according to Feynman, the Universe is constantly spitting into an infinite number of branches, as a result of the
interactions between its components. As a consequence,
an inside observer knows which branch the world has taken, so that
he can follow the track of his past. However, in 1981 Feynman says of the many-worlds picture: \textit{It's possible, but I am not very happy with it}.

\section{Conclusion}

Physics enables everything and underlies everything, but it does not mean it determines everything \cite{ellistopdown}. Perhaps the world is much more complicated than can be explained only by science. Thus, we should trust our most immediate experiences, new conceptual and technical challenges. In fact, they are the basis and guides for being capable of thinking and acting. Based on these, we could critically examine and take care of metaphysical claims and our ultimate questions like: does the world covered by science describe the whole reality? What about underlying assumptions and axioms in cosmology? What about all the philosophical issues related to "the interpretation of quantum mechanics," "measurement Problem," and the ontological character of quantum states? Why does the Universe have such a unique structure that life can exist?

\bibliographystyle{ws-procs961x669}
\bibliography{ws-pro-sample.bib}

\end{document}